\documentclass[aps,prd,twocolumn,showpacs,superscriptaddress,preprintnumbers,floatfix,amsmath,amssymb,eufrak,table]{revtex4-1}

\usepackage{graphicx,fancybox}
\usepackage{amsmath,amssymb,mathrsfs,slashed}
\usepackage[title]{appendix}
\usepackage[colorlinks=true, pdfstartview=FitV, linkcolor=red, citecolor=blue, urlcolor=blue]{hyperref}
\usepackage{color}
\usepackage{soul}
\usepackage{array}
\usepackage{url}
\usepackage{bm}
\usepackage{dcolumn}
\usepackage{bm}
\usepackage{braket,xfrac,lineno}
\usepackage[utf8]{inputenc}
\hypersetup{
colorlinks=true,
pdfstartview=FitV, 
linkcolor=blue,
citecolor=red,
urlcolor=blue
}

\newcommand \ra  {\rightarrow}

\newcommand \A {\alpha}

\newcommand \lc {\langle}
\newcommand \rc {\rangle}

\newcommand \bvec{\left( \begin{array}{c} }
\newcommand \evec{\end{array} \right)}
\newcommand \bmat{\left( \begin{array}}
\newcommand \emat{\end{array} \right)}

\newcommand \bea{\begin{eqnarray} }
\newcommand \eea{\end{eqnarray} } 
\newcommand \nn {\nonumber}
\newcommand {\be} {\begin{equation}}
\newcommand {\ee} {\end{equation}}

\newcommand {\mbx} {\mbox{}}

\newcommand {\psibar} {\bar{\psi}}
\newcommand{\qhat}{$\hat{q}$ }

\begin{document}
\title{Jet transport coefficient $\hat{q}$ in lattice QCD}

\author{Amit~Kumar} 
\email[]{amit.kumar3@mail.mcgill.ca}
\affiliation{Department of Physics and Astronomy, Wayne State University, Detroit, MI 48201, USA}
\affiliation{Department of Physics, McGill University, Montreal, QC H3A-2T8, Canada}

\author{Abhijit~Majumder}
\email[]{majumder@wayne.edu}
\affiliation{Department of Physics and Astronomy, Wayne State University, Detroit, MI 48201, USA}

\author{Johannes~Heinrich~Weber}
\email[]{johannes.weber@physik.hu-berlin.de}
\affiliation{
Department of Computational Mathematics, Science and Engineering \& Department of Physics and
Astronomy, Michigan State University, East Lansing, MI 48824, USA}
\affiliation{Institut f\"ur Physik, Humboldt-Universit\"at zu Berlin \& IRIS Adlershof, D-12489 Berlin, Germany}

\date{\today}
\preprint{HU-EP-20/26-RTG}

\begin{abstract} 
We present the first calculation of the jet transport coefficient $\hat{q}$ in quenched and (2+1)-flavor QCD on a 4-D Euclidean lattice. 
The light-like propagation of an energetic parton is factorized from the mean square gain in momentum transverse to the direction of propagation, 
which is expressed in terms of the thermal field-strength field-strength correlator. The leading-twist term in its operator product expansion is calculated on the lattice.
Continuum extrapolated quenched results, and full QCD estimates based on un-renormalized lattice data, over multiple lattice sizes, are compared with (non) perturbative calculations and phenomenological extractions of \qhat\!\!.  The lattice data for \qhat\!\! show a temperature dependence similar to the entropy density. Within uncertainties, these are consistent with phenomenological extractions, contrary to calculations using perturbation theory. 
\end{abstract}

\maketitle

\section{Introduction}
The study of hot and dense QCD matter, produced in relativistic 
heavy-ion collisions, using high transverse momentum ($p_T$) jets, currently 
boasts an almost established phenomenology~\cite{Majumder:2010qh, 
Cao:2020wlm,Burke:2013yra}. 
The experimental data on various aspects of jet modification is also 
extensive 
~\cite{Adler:2006bw,Adare:2010sp,Adams:2006tj,Adler:2002tq, 
Chatrchyan:2013exa,Acharya:2017goa,Cole:2011zz,Abelev:2013kqa,
Chatrchyan:2014ava,CMS:2011kra,Aad:2014bxa}. 
Almost all of the evidence points to the formation of a quark-gluon 
plasma (QGP), a state of matter where the QCD color charge is 
deconfined over distances larger than the size of a 
proton~\cite{Shuryak:1980tp, Shuryak:1978ij}. 
Chiral symmetry -- spontaneously broken in a hadron gas -- 
is restored during the transition to the QGP, which is a smooth crossover 
at zero baryon density centered around the pseudo-critical temperature 
$T_{pc}=156.5(1.5)$~MeV~\cite{HotQCD:2018pds,Borsanyi:2020fev} 
(for three physical light quark flavors in the sea).
Jets are expected to undergo considerable modification within the QGP compared 
to confined nuclear matter~\cite*{Gyulassy:1993hr}.

While a lot of the theoretical development of jet quenching has been focused on modifications to the parton shower, 
considerably less work has been carried out on the study of the interaction between a parton in the jet with the QGP itself.
Most current calculations either model the QGP as a set of slowly moving (or static) heavy scattering centers~\cite{Gyulassy:1993hr,Baier:1994bd,Wang:1994fx,Baier:1996sk}, or in terms of Hard-Thermal Loop (HTL) effective theory~\cite{Frenkel:1989br,Braaten:1989kk,Braaten:1989mz,Arnold:2002ja}. 
Regardless of the model, a description of transverse momentum exchange between the medium and a jet parton
can be encapsulated within the transport coefficient~\cite{Baier:2002tc}
\bea
\hat{q} =     \frac{\sum_{i=1}^{N_{events}} \sum_{j=1}^{N^i (L)} [k_\perp^{i,j}]^2}{ N_{events} \times L}.
\eea
The meaning of the above equation is that given a path through a medium with a pre-determined density profile, 
a single parton may scatter $N^i (L)$ times while traversing a distance $L < v \tau_i$ in event $i$ 
($\tau_i$ is the lifetime of the parton which travels at a speed $v$). 
In each scattering ($j$), it exchanges transverse momentum $k_\perp^{i,j}$. In this paper, we will only focus on momentum exchanges transverse to the direction of the jet parton, as these tend to have a dominant effect on the amount of energy lost via bremsstrahlung from the parton~\cite{Baier:1996sk,Baier:1996kr}.

In heavy-ion collisions, the density will vary with location,
and thus, one necessarily averages over a non-uniform profile, which fluctuates from event to event. 
Several successful fluid dynamical simulations, which compare to RHIC and LHC data~\cite{Song:2010mg,Shen:2011eg}, assuming small density gradients, have used an equation of state calculated in lattice QCD~\cite{Huovinen:2009yb} as an input. Unlike the dynamical medium in a heavy-ion collision, 
lattice simulations assume static media in thermal equilibrium. 
The use of lattice QCD input in fluid-dynamical simulations is predicated on the ability to reliably coarse grain the system into space-time unit cells, over which intrinsic quantities, e.g., temperature ($T$), 
entropy density ($s$), pressure ($P$), remain approximately constant.

The calculations presented in this paper are an extension of the above principle: Calculations of $\hat{q}$ in 
the static medium of lattice QCD will be compared with phenomenological estimations, where jets are propagated through a QGP fluid dynamical 
simulation.  These QGP simulations yield the space-time profiles for intrinsic quantities, e.g. $T(\vec{r},t)$, $s(\vec{r},t)$ etc., and the local \qhat is calculated from these, using dimensional parametrization or perturbative techniques, typically with an overall normalization that can be varied to fit experimental data. Thus, the parton propagating through this dynamical medium experiences a varying \qhat\!. Once the overall normalization is determined, one reports the \qhat (or some dimensionless ratio involving \qhat) as a function of the temperature. 

In this paper, we will compute the dimensionless ratio $\sfrac{\hat{q}}{T^3}$~\cite{Burke:2013yra}, directly from lattice gauge theory and compare with calculations of the same ratio using other models of the QGP, and parametrized extractions from comparisons with experimental data. The paper is organized as follows: In Sec.~\ref{sec: qhat-definition}, we outline the basic process of a single parton scattering off the glue field in a QGP, define \qhat and relate it to a series of local operator products with increasing number of covariant derivatives, suppressed by increasing powers of the energy of the parton. 
In Sec.~\ref{sec:lattice-details}, focusing only on the leading operator product, in the limit of a very energetic parton, we present details of the calculation of this operator product in both quenched and full QCD simulations. Results of our calculations, compared to those from both model calculations and phenomenological extractions from data, are presented in Sec.~\ref{sec:results}. A summary and outlook for future calculations is presented in Sec.~\ref{sec:conclusions}. Numerical tables derived from the lattice ensembles used, as well as perturbatively calculated correction factors are discussed in the appendices. 
\section{ Jet transport coefficient \qhat }
\label{sec: qhat-definition}

The transport coefficient \qhat is the leading jet transport coefficient that characterizes the rate of medium-induced radiative energy loss of the hard parton traversing the QGP. 
A strategy to compute this coefficient from first principles within the framework of lattice gauge theory was first proposed in Ref. \cite{Majumder:2012sh}. In this section, we briefly describe the methodology and formulate \qhat in terms of a series of local operators that can be computed on pure gluonic plasma and  QCD plasma.

\subsection{Leading Order expression }
We consider a hard  parton with high energy $E$ and virtuality $Q$ such that $E \!\gg\! Q\! \gg \!\mu_D$, the Debye mass in the medium. 
The choice of a large $E,Q$ leads to a diminished coupling $\alpha_S(Q)$ with the medium, due to asymptotic freedom~\cite{Politzer:1973fx,Gross:1973id}. 
As a result, interactions between the hard parton and a medium of limited extent will be dominated by one-gluon exchange (OGE); i.e. $N^i=1$ for all events. 

\begin{figure}[h!]
  \centering
    \includegraphics[width=0.5\linewidth]{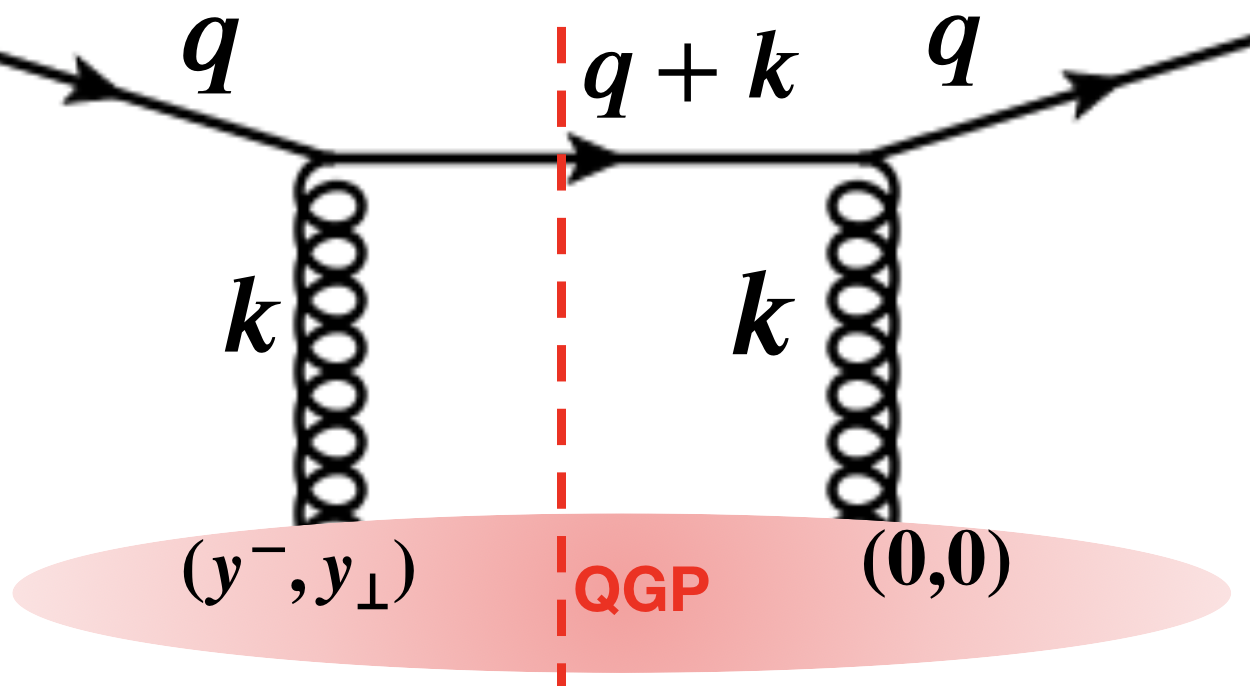}
  \caption{A forward scattering diagram for the hard quark undergoing a single scattering off the gluon field  in the plasma. The vertical dashed line represents the cut-line.}
  \label{fig:ForwardScatteringDiagram}
\end{figure}

In light-cone coordinates, the incoming quark, traveling in the $-z$ direction, has two non-zero components, 
$q^+ = \sfrac{(q^0 + q^3)}{\sqrt{2}} \ll q^- = \sfrac{(q^0 - q^3)}{\sqrt{2}}$.
The quark undergoes a single scattering off the gluon field in the medium  
and gains transverse momentum $k_\perp$ (Fig. \ref{fig:ForwardScatteringDiagram}).  In this frame, the momentum of the quark changes from, 
\bea
q_i \equiv [q^+,q^-,0,0]  \ra  q_f \equiv [q^+ + \vec{k}_\perp^2/(2q^-), q^- , \vec{k}_\perp].
\eea
The momentum scaling of incoming quark and exchanged gluons are 
\begin{equation}
    \begin{split}
        q_{i} & \equiv (Q^{2}/2q^{-},q^{-}, 0_{\perp} ) \sim (\lambda^2,1,0)q^{-}, \\
        k & \equiv (k^{+}, k^{-},k_{\perp}) \sim (\lambda^{2},\lambda^{2},\lambda)q^{-},
    \end{split}
\end{equation}
where $\lambda \ll 1$. 
The matrix element for this process is given as 
\begin{equation}
 \mathcal{M} \! \!= \! \lc q_f | \otimes \lc X | \! \!\int \limits_0^{T_I}\! d t d^3 x g \psibar
(x) \gamma^\mu t^a A^a_\mu (x) \psi(x) | n \rc  \otimes | q_i \rc, 
\end{equation}
where $|n\rc$ and $|X\rc$ represent the initial and final state of the medium, respectively. The factors $\psi(x)$, $\psibar(x)[=\!\psi^\dag \gamma^0]$ and $A^a_{\mu}(x)$ represent the 
quark and gluon wave functions (and complex conjugate), with coupling $g$. The spatial integrations are limited within a volume $V=L^3$ and the time of interaction ranges from $0$ to $T_I=\sfrac{L}{c}$ (we use particle physics units with $\hbar, c=1$). 
Replacing the average over events, with an average over all initial states $|n\rc$ (energy $E_n$) of the medium, weighted by a Boltzmann factor, with $\beta=1/T$ the inverse temperature and $Z$ the  partition function of the thermal medium, we obtain, 
\bea
\mbx\!\!\!\!\hat{q} = \sum_{n} \frac{e^{-\beta E_n}}{Z T_I} \!\! \int \!\!d^4 k k_\perp^2 
\times \frac{d^{4}W(k)^{n,X} } {d^{4}k},
\label{eq:qhatDefinitionBoltzmann}
\eea
%
where  
\bea
W^{n,X} = \frac{|\mathcal{M}|^{2}}{2N_{c}}
\eea
represents the scattering probability, for a quark in one of 2 spins and $N_c$ color states.
After performing spin sum and colored average, the differential decay rate is given as (assuming $\sum_X |X\rc\lc X| = 1$):
\begin{equation}
\begin{split}
 \frac{d^{4}W^{n,X}(k)}{d^4k}  
  = & \frac{Disc}{2\pi i}\left[ \frac{g^2}{(2\pi)^4V N_{c}}\int d^4xd^4y  \frac{ e^{-ik(x-y)}}{2E(q+k)^2} \right. \\
 & \left. \times \bra{n}  \mathrm{Tr} \left[ \slashed{q}   \slashed{A}(x) (\slashed{q} + \slashed{k}) \slashed{A}(y)     \right] \ket{n} \right], 
\end{split}
\end{equation}
where $N_{c}$ is the number of colors.

Following standard methods outlined in Ref.~\cite{Majumder:2012sh}, where factors of $k_\perp$ are turned into partial derivatives in $y_\perp$ and $y^-$, we obtain the following well known expression for $\hat{q}$ ,
\bea
\hat{q} &= &  c_{0}  \int \frac{ dy^{-} d^{2} y_{\perp} } {(2\pi)^3}  d^{2} k_{\perp}  e^{ -i  \frac{\vec{k}^{2}_{\perp}}{2q^{-}} y^{-} +i\vec{k}_{\perp}.\vec{y}_{\perp} }  \nonumber \\
& \times& \sum _{n}  \bra{n} \frac{e^{-\beta E_{n}}}{Z}  \mathrm{Tr}[F^{+ j}(0) F^{+}_{j}(y^{-},y_{\perp})] \ket{n}, \label{eq:OGE-qhat}
\eea
 where $c_{0}\!=\!16 \pi \alpha_s \sqrt{2} \sfrac{C_R}{( N_{c}^2 -1 )}$, 
 $C_R$ (for a quark $C_{R}= C_F = \sfrac{( N_C^2 -1 )}{(2N_C)}$) is the representation specific Casimir, $\alpha_s$ is the strong coupling constant  at the vertex between the hard quark and the glue field, $F^{ \mu\nu} = t^{ a} F^{ a\mu\nu}$ is the bare gauge field-strength tensor.
Here and hereafter the index $j=1,2$ runs over transverse directions. 
 
Computing the thermal and vacuum expectation value of the non-perturbative operator $ F^{+ j}(0) F^{+}_{j}(y^{-},y_{\perp})$ is challenging due to the \emph{near} light-cone separation between the two field-strength tensors. The separation is slightly space-like $y^2 = - y_\perp^2 < 0$, similar to the case of a parton distribution function (PDF)~\cite{Ji:2013dva}. 
Beyond this method, there have been other efforts based on 
a \mbox{3-D} Euclidean lattice approach~\cite{Panero:2013pla,Moore:2021jwe}, as well as in classical lattice theory~\cite{Laine:2013apa}. 
Another non-perturbative pure-glue calculation of $\hat{q}$ employed a stochastic vacuum model~\cite{Antonov:2007sh} with inputs obtained from lattice simulations. 
However, the current framework remains the sole exploration of $\hat{q}$ in \mbox{4-D}, first-principles quantum lattice simulations. 

 \subsection{Analytic continuation in deep-Euclidean region}

To recast $\hat{q}$ in terms of a series of local operators amenable to a lattice calculation, we apply a method of dispersion as outlined in Ref. \cite{Majumder:2012sh}. In this approach, a generalized coefficient is defined as, 
\begin{equation}
\begin{split}
\mbx\!\! \hat{Q}(q^{+}) \! = \! c_{0} \! \!\! \int \!\! \frac{ d^{4}y d^{4}k e^{iky} }{(2\pi)^4}   
 \frac{ 2q^- \lc  \mathrm{Tr} [ F^{ + j} (0) F^{+}_{j} \!(y) ] \rc }{(q+k)^2 + i\epsilon },
\end{split}
\label{Eq:GeneralizedQHatDef}
\end{equation} 
where  
\bea
\lc \dots \rc \equiv \sum_{n} \bra{n} \ldots \ket{n} \frac{e^{-\beta E_{n}}} {Z}.
\eea
The object $\hat{Q}(q^{+})$ has a branch cut in a region where $q^{+} \sim T \ll q^{-}$ corresponding to the quark propagator with momentum $q+k$ going on mass shell (Fig. \ref{fig:ForwardScatteringDiagram}). In this region, the incoming hard quark is light-like, i.e. 
$q^{2}=2q^{+}q^{-} \approx 0$, 
and the discontinuity of $\hat{Q}(q^{+})$ is related to the physical $\hat{q}$ as  
\begin{equation}
\left. \frac{ Disc[\hat{Q}(q^{+})] }{2\pi i} \large \right | _{ \hspace{1mm } \mathrm{at}  \hspace{1mm} q^{+} \sim T } = \hat{q}.
\label{eq:qhatDiscontinuity}
\end{equation}

In addition to the thermal discontinuity, $\hat{Q}(q^{+})$ also has an additional vacuum discontinuity in the region $q^{+} \in (0, \infty )$ due to real hard gluon emission processes. In this region, the incoming hard quark is time-like.
If instead, one takes $q^{+} \ll 0$, e.g. $q^{+}=-q^{-}$, the incoming quark becomes space-like and there is no discontinuity on the real axis of $q^+$. In this deep space-like region, the quark propagator can be expanded as follows:
\bea
\begin{split}
\mbx \! \!\!\!\!\! \frac{1}{(q+k)^2}  \! &\simeq   \! \frac{-1}{\! 2q^{-} ( q^{-} \! - \! (k^{+} \! - \! k^{-} )  )} \\
 \! &=\!  \frac{-1}{2 (q^{ -})^2} \!  \sum^{\infty} _{n=0} \! \left(  \!\! \frac{\sqrt{2}k_{3}}{q^{-}} \!\! \right)^{\!\! n}\!\!\!.
\end{split}
\label{eq:QuarkPropagatorExpansion}
\eea

Using integration by parts, the factor of exchanged gluon momentum $k_{3}$ [Eq.~\eqref{eq:QuarkPropagatorExpansion}] 
is replaced with the regular spatial derivative $\partial_{3}$ acting on the field-strength $F^{+}_{j}(y)$ [Eq.~\eqref{Eq:GeneralizedQHatDef}].  A set of higher order contributions from gluon scattering diagrams can be added to promote the regular derivative to a covariant derivative $D_3$ (in the adjoint representation). With all factors of $k$ removed from the integrand [Eq. \eqref{Eq:GeneralizedQHatDef}], except for the phase factor, $k$ can be integrated out ($\int d^4k e^{iky}$) to yield $\delta^{4}(y)$, setting $y$ to the origin. This yields 
$\hat{Q} (q^+ = -q^-)\,[\equiv\hat{Q}|_{q^+ = -q^-}]$ as,
\bea
\left. \mbx\!\!\!\!\!\!\!\!\hat{Q}\right|_{\mbx_{q^+ = -q^-}} \!\!\!\!\!\!\!\!\!\!\!\!\!=   c_0   
 \lc  \mathrm{Tr} [F^{+ j}(0) \sum^{\infty}\limits_{n=0}  
 \left(   \frac{i\sqrt{2}D_{3
 } }{q^{-}}\right)^{n}\!\! F^{+}_{j}(0) ]  \rc / q^{-} ; \label{Qhat-expansion}
\eea 
each term in the series is a local gauge-invariant operator.

To relate $\hat{Q}(q^{+}\!\!=\!\!-q^{-})$ to the physical $\hat{q}$, consider the following contour integral in the $q^{+}$ complex-plane:
\bea
I_{1} = \oint \frac{dq^{+}}{2\pi i} \frac{\hat{Q}(q^{+})}{(q^{+}+q^{-})} = \hat{Q}(q^{+}=-q^{-}),
\label{eq:ContourI1}
\eea
where the contour is taken as a small anti-clockwise circle centered around point $q^{+}\!\!=\!\!-q^{-}$, with a radius small enough to exclude regions where $\hat{Q}(q^{+})$ may have discontinuities. 
Alternatively, the integral can be evaluated by analytically deforming the contour over the branch cut of $\hat{Q}(q^+)$ for $q^{+} \in (-  T_{1}, \infty)$ and obtaining Eq.~\eqref{eq:ContourI1} as:
\begin{equation}
\begin{split}
\mbx\!\!\!\! \hat{Q}(q^{\! +} \!\!=\! - q^{\! -\!}) \! =\!\! \!\! \int \limits ^{T_2} _{-T_1} 
\!\!\! \frac{dq^{+}}{2\pi i}  \frac{\! Disc[\hat{Q}(\!q^{ +}\!)]\!}{\!(q^{ +} \! +  q^{-}) }   
 \! + \!\! \int \limits^{\infty}_{0} \!\frac{\! dq^{\! +}}{\! 2\pi i \!} \frac{Disc[ \hat{Q}(q^{ +}\!)]}{(q^{+} \! + q^{-})}.
\end{split}\!\!\!
\label{eq:ContourEquationThermalPlusVacuum}
\end{equation}
The limits $-T_{1}$ and $T_{2}$ in the first integral represent lower and upper bounds of $q^{+}[=k^+ \!\!+ \sfrac{k_\perp^2}{(2q^- \!\!+ 2k^-)}]$, beyond which the thermal discontinuity in $\hat{Q}(q^{+})$ on the real axis of $q^{+}$ is zero. In this region, the hard incoming quark is close to on-shell, i.e. $q^2=2q^{+}q^{-} \approx 0$ and undergoes scattering with the medium.  
 The second integral in Eq.~\eqref{eq:ContourEquationThermalPlusVacuum} corresponds to the contributions from vacuum-like processes, where the time-like hard quark with momentum $q^{+} \in (0, \infty)$ undergoes vacuum-like splitting. Hence, the second integral is temperature independent.

Using Eqs.~(\ref{eq:qhatDiscontinuity}-\ref{eq:ContourEquationThermalPlusVacuum}), we obtain (suppressing $y^-\!=\!y_\perp\!=\!0$),
\begin{equation}
\mbx\!\!\!\!\frac{\hat{q}}{T^3} = \frac{ c_{0} \!\!
\sum\limits^{\infty}_{n=0} \!\left(\!\frac{T}{q^{-}}\!\right)^{\!\!2n}\!\!\!
\left\lc \!\!\frac{1}{T^4} \mathrm{Tr}\left[ F^{+j} 
%
 \Delta^{2n} \! F^{+}_{j} \!  \right]\!\right\rc_{(\mathrm{T-V})}
}{ \sfrac{(T_{1} + T_{2})}{T}} ,
\label{eq:qhatLatticeEquation}
\end{equation}
where the subscript $(\mathrm{T-V})$ represents the vacuum subtracted expectation value and 
$T_{1}+T_{2} \simeq 2\sqrt{2}T$ represents a width of the thermal discontinuity 
of $\hat{Q}(q^{+})$. 
The width of the discontinuity  in Cartesian coordinates will always be very close to $2T$~\footnote{As can be calculated in HTL perturbation theory}. The extra factor of $\sqrt{2}$ is due to the choice of light cone coordinates. Minor shifts in this estimate may depend on details of how the medium itself is treated, on the nature of the parton, or on the loop order of the interaction.
We abbreviate the differential operator as $\Delta \equiv \sfrac{i\sqrt{2}D_3}{T}$.
Only even powers of $\Delta$ contribute in Eq.~\eqref{eq:qhatLatticeEquation}, 
since $\left\lc \mathrm{Tr}\left[ F^{+j} \Delta^{2n+1} \! F^{+}_{j} \!  \right]\right\rc$ 
would not be invariant under either parity or time-reflection, and thus evaluates to zero. 
The above expression for the transport coefficient $\hat{q}$  contains several features: 
Each term in the series is local, allowing for their computation on the lattice. 
The successive terms in the series are suppressed by the hard scale $q^{-}$, 
and hence, computing only the first few terms may be sufficient. 
In the limit $q^- \ra \infty$ only the leading-twist term contributes, namely, the first term in the series [Eq.~\eqref{eq:qhatLatticeEquation}].

To compute the local operators [Eq.~\eqref{eq:qhatLatticeEquation}] at finite temperature, we perform Wick's rotation  
\bea
x^{0} \!\rightarrow\! -ix^4, A^{0} \!\rightarrow\! iA^{4}\! 
\Longrightarrow \!F^{0j}\! \rightarrow \!iF^{4j}.
\eea
For a quark in the limit of $q^- \ra \infty$, \qhat  reduces to 
\begin{equation}
\frac{\hat{q}}{T^3} = \frac{4\pi \alpha_s }{N_C T^4}  \lc F^{+ j} F^{+}_j \rc_{\mathrm{T-V}} 
\end{equation}
In the case of a quenched plasma of gluons, the expectation value $\lc F^{+ j} F^{+}_j \rc_{\mathrm{T-V}} $ is related to the entropy density $s$ via the following relation: 
\begin{equation}
\frac{1}{T^4}  \lc F^{+ j} F^{+}_j \rc_{\mathrm{T-V}} 
= \frac{1}{2} \frac{s}{T^3}.
\end{equation}
Hence, we obtain a direct relation between \qhat and $s$: 
\begin{equation}
\hat{q} = \frac{2\pi \A_s}{N_C} s. \label{q-hat-s}
\end{equation}

Since $s$ is a genuine physical observable (protected by Ward identities) that does not require renormalization (in the continuum), the renormalization of $\A_s$ (in $\overline{MS}$ scheme) introduces an unavoidable scheme dependence of \qhat. 
Note, the above relation [Eq.~\eqref{q-hat-s}] holds for the case of infinite energy quark traversing pure SU(3) plasma.

\subsection{ Analytic relation in a weakly coupled theory}
\label{subs:WeakCouplingRelation}

The $\hat{q}$ relation derived in [Eq.~\eqref{q-hat-s}]  relating a dynamical quantity $\hat{q}$ to a static quantity $s$, may seem hard to believe at first, but 
one can show that this indeed holds in the limit of running coupling for a very high energy parton. 
In this subsection, we will demonstrate this for a weakly-coupled quenched QGP where analytical expressions exist for both sides of the equation. To obtain a $\hat{q}$ that is well defined in the limit of $q^- \ra \infty$ or $E \ra \infty$, we use the expression including running coupling derived by Arnold and Xiao~\cite{Arnold:2008vd}, in the limit that $E \gg T \sim m_D$, where $m_D$ is the Debye mass, 
\bea
\hat{q} = C_F  \Xi_b \mathcal{I}_+ (\Lambda)    g^2(\Lambda) g^2(m_D) \frac{T^3}{\pi^2} . \label{quark-qhat-HTL-0}
\eea
In the quenched limit for the plasma, we only include the sum over spin degrees of freedom times the trace normalization for gluons, where $\Xi_b = 2 C_A = 6$. The factor $\mathcal{I}_+ (\Lambda)$ contains large logarithms which depend on the hard scale $\Lambda$, which we assume to be $\sqrt{ \mathcal{N} E T}$, where $2 \lesssim \mathcal{N} \lesssim 6$. Thus, we have 
\bea
\mathcal{I}_+ &=& \frac{\zeta(3)}{2\pi} \ln \left( \frac{\Lambda}{m_D} \right) + \Delta \mathcal{I}_+.
\eea
The correction term,
\bea
\Delta \mathcal{I}_+ &=& \left[ \frac{ \left( \zeta(2) - \zeta(3) \right)}{2\pi}  \right] \left[ \ln \left( \frac{T}{m_D} \right) \right. \nn \\ 
&+& \left. 1/2 -\gamma_E + \ln(2) \right] - 0.386/(2\pi), 
\eea
does not contain any large logarithms involving the hard scale $\Lambda$ and can be neglected compared to the leading term in $\mathcal{I}_+$.

We can now combine the large logarithm in $\mathcal{I}_+$ along with that in $g^2 $ to obtain 
\bea
\lim_{\Lambda \ra \infty} g^2(\Lambda) g^2(m_D) \ln \left( \frac{ \Lambda }{m_D} \right)  \simeq \frac{g^2(m_D)}{ -2 \beta_0},
\eea
where 
\bea
\beta_0 = - \frac{11C_A - 4 N_f T_F}{48\pi^2} = -\frac{ 11 C_A}{48 \pi^2} \,\,\, \{ {\rm for} \,\, N_f = 0   \},
\eea
and $T_{F}=\sfrac{1}{2}$.
 Substituting the above in the expression for $\hat{q}$ in Eq.~\eqref{quark-qhat-HTL-0}, we obtain 
\bea
\hat{q} &=& C_F \Xi_b \frac{\zeta(3) T^3}{4\pi} \frac{48}{11C_A} g^2 (m_D)  \nn \\
&=& \frac{N_C^2 - 1}{N_C} \alpha_s(m_D)  T^3 \left\{  \zeta(3) \frac{48}{11} \right\} ,
\eea
where we have separated the obvious factors of color, coupling and temperature from the residual numerical factor in curly brackets. 
Using the Ramanujan series expansion for Apery's constant $\zeta(3) \simeq \sfrac{7\pi^3}{180} $, we obtain 
\bea
\hat{q} = \frac{N_C^2 - 1}{N_C} \alpha_s(m_D)  T^3  \left\{  \frac{ 8 \pi^3}{45 } \frac{ 21}{22}        \right\} .  \label{final-qhat-LO}
\eea

The entropy density of a pure non-interacting (massless) gluon gas is given as,
\bea
s =  (N_C^2 - 1) T^3 \pi^2 \frac{4}{45}.
\eea
Substituting the above into Eq.~\eqref{q-hat-s} and separating factors of color, coupling and temperature from the residual numerical factor, we obtain
\bea
\hat{q} = \frac{N_C^2 - 1}{N_C} \alpha_s  T^3 \left\{   \frac{8 \pi^3 }{45}    \right\}.  \label{final-qhat-s}
\eea
Evaluating $\alpha_s$ at $m_D$, one notes that the two methods to obtain $\hat{q}$ are within 5\% of each other. We point out, that in the expression for the entropy density above, we have neglected any effect of dynamically generated thermal mass, while dynamically generated screening effects are included in the expression in Eq.~\eqref{final-qhat-LO}. The inclusion of these effects will reduce the entropy density and bring Eq.~\eqref{final-qhat-s} even closer to Eq.~\eqref{final-qhat-LO}.
\section{Computing \qhat on 4D Lattice}
\label{sec:lattice-details}

After having confirmed the  veracity of the formalism introduced in Ref. \cite{Majumder:2012sh}  for the case of an energetic parton traversing a weakly-coupled quenched plasma, we proceed to more realistic plasmas simulated on a lattice with standard periodic boundary conditions. 
As we have seen earlier [Eq.~\eqref{q-hat-s}], the leading-twist operator can be related for a pure glue plasma to the energy momentum tensor (EMT). 
The EMT's non-singlet components, which are in a nonet representation in the 
continuum, split into a triplet and a sextet in the discretized theory that 
require multiplicative renormalization~\cite{Caracciolo:1989pt}.
Operators mixing magnetic and electric field strengths that are included 
in Eq.~\eqref{eq:qhatLatticeEquation} are related to the sextet representation of the energy-momentum tensor, and hence vanish on ensembles 
with our chosen boundary conditions, i.e. in the rest frame.
We have studied the first three non-zero operators in the $\hat{q}$ series, 
\bea 
\hat{O}_n=\frac{\mathrm{Tr}[ F_{3j} \Delta^{2n} F_{3j} \!-\! F_{4j} \Delta^{2n} F_{4j}]}{T^4}
\label{eq:On}
\eea
~(summed over $j$; $n=0,1,2$).
The field-strength $F_{\mu\nu}$ is discretized  
via clover-leaf operators projected to 
anti-Hermitian traceless matrices, 
\bea
\mbx\!\!\!\!\!i g_0\mathcal{F}_{\mu\nu}(x) &=& \frac{\mathcal{P} Q_{\mu\nu}(x)}{a_L^2 } 
= i g_0 F_{\mu\nu} + \mathcal{O}(a_L^2), 
\eea 
where  
\bea
Q_{\mu\nu}= \frac{1}{4}[U_{\mu,\nu} + U_{\nu,-\mu} + U_{-\mu,-\nu}+ U_{-\nu,\mu}]
\label{eq:clover}
\eea
with  $U_{\mu,\nu}$ being the plaquette with lattice spacing $a_{L}$ in plane $\mu$-$\nu$ 
and 
\bea
\mathcal{P} Q =\frac{1}{2} \left(Q-Q^\dagger -\frac{1}{N_c}\mathrm{Tr}[Q-Q^\dagger] \right).
\eea
%
%

\subsection{Gauge ensembles and lattice setup}

In this subsection, we discuss the parameters used in generating gauge ensembles for pure SU(3) and (2+1)-flavor QCD lattices.
We have generated $n_{\tau}\times n^3_{\sigma}$ lattices at  finite temperature $T\!>\!0$  with aspect ratio $\sfrac{n_\sigma\!}{n_\tau}\!\!=\!4$ for $n_\tau\!\!=\!4,6$ and $8$, where  $T\!\!=\!\sfrac{1}{(a_L n_\tau)}$, 
and the corresponding vacuum $T\!\!=\!0$ lattices with $n_{\tau}\!\!=\!\!n_{\sigma}$ 
using the MILC code~\cite{MILCcodePackage} 
and the USQCD software stack~\cite{SciDACcodePackage}. 

 In the presented lattice calculations, the unquenched lattices were generated using the Rational Hybrid Monte Carlo algorithm (RHMC)~\cite{Clark:2004cp} 
with highly improved staggered quark~(HISQ) action~\cite{Follana_2007} and 
tree-level Symanzik gauge action~\cite{Bazavov:2014pvz,Bazavov:2018wmo} 
for (2+1)-flavor QCD. 
The leading cutoff effects are $\mathcal{O}(a^4)$ and 
$\mathcal{O}(g^2_{0}a^2)$. 
We employed tuned input parameters (bare lattice coupling $\beta_0=10/g_0^2$, 
and bare quark masses), and use the $r_1$ lattice scale following 
Refs.~\cite{Bazavov:2014pvz,Bazavov:2016uvm,Bazavov:2017dsy,Bazavov:2018wmo} by the HotQCD and 
TUMQCD collaborations. 
This setup has a physical strange and two degenerate light quarks with 
$m_{l}=m_{s}/20$ corresponding to a pion mass of about 160 MeV in the 
continuum limit.
We summarize the gauge ensembles in Tables \ref{table:FullQCDNt4},  \ref{table:FullQCDNt6} 
and \ref{table:FullQCDNt8} in  Appendix~\ref{sec:GaugeEnsemble}.

We have generated pure SU(3) gauge ensembles via the heat-bath algorithm using Wilson gauge action~\cite{Wilson:1974sk} with $\beta_{0}\!=\!6/g^{2}_{0}$ and leading 
cutoff effects $\mathcal{O}(a^2)$.
%
We summarize the gauge ensembles in Tables \ref{table:PureSU3Nt4}, \ref{table:PureSU3Nt6}  and \ref{table:PureSU3Nt8} 
in Appendix~\ref{sec:GaugeEnsemble}, where we also discuss the scale setting.

\subsection{Temperature dependence of bare operators } 

\begin{figure}[h!]
  \centering
    \includegraphics[width=0.8\linewidth]{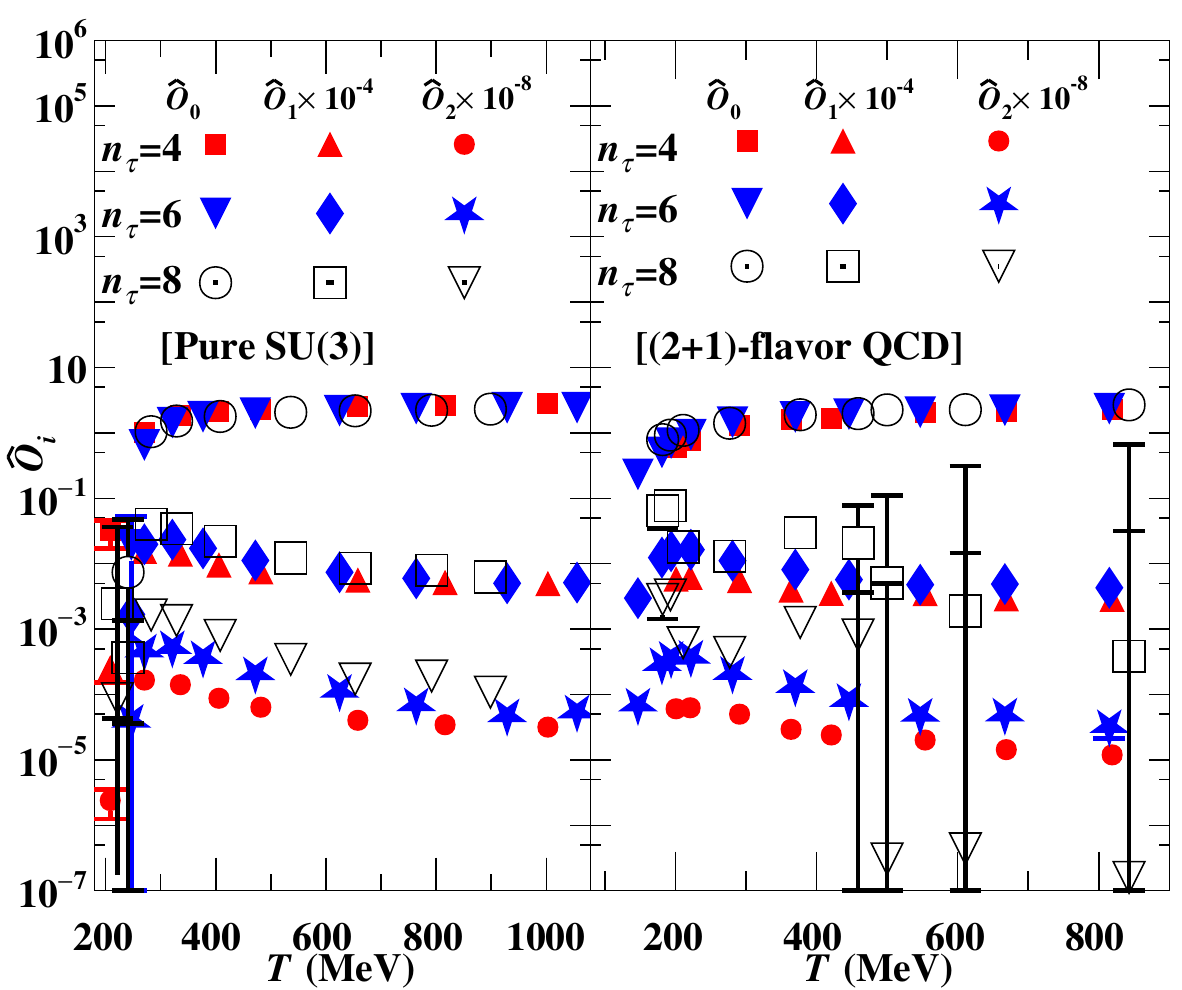}
  \caption{
  Temperature dependence of vacuum-subtracted field-strength correlators 
on quenched and unquenched SU(3) lattices. The operators are unrenormalized and have been computed for lattice sizes $n_\tau=4,6$ and $8$. }
  \label{fig:LatticeFFcorrelatorsResult}
\end{figure}

We present in Fig. \ref{fig:LatticeFFcorrelatorsResult} the expectation values of 
the bare field strength operators $\hat{O}_{n}$ [Eq.~\eqref{eq:On}] for all 
ensembles ($n_\tau=8,~6$~and $4$) in pure gauge theory or (2+1)-flavor QCD. 
The vacuum contributions have been subtracted while computing the temperature dependence; for the leading-twist operator, $\hat{O}_0$, the vacuum contribution vanishes within the statistical error as naively expected (In the one gluon exchange approximation, the vacuum contribution corresponds to the difference of the transverse electric and magnetic field squared of a radiated on-shell gluon, which is identically zero).

The operator $\hat{O}_0$ exhibits a rapid transition near the temperature $T\in (150,250)$ MeV for full QCD case and $T \in (250,350)$ MeV for pure SU(3) gauge theory.
The operators with derivatives $\Delta$ are scaled by the factor $10^{-4}$ and $10^{-8}$ to illustrate the ordering of operators as an overall factor of $(T/q^{-})^{2n}$ appears in the $\hat{q}$ expression [Eq. \eqref{eq:qhatLatticeEquation}]; it corresponds to a hard parton, i.e. $T \sim 1$ GeV and $q^{-} \!\sim 100$ GeV. Looking at the operators $\hat{O}_{1}$ and $\hat{O}_{2}$ for pure SU(3) case where the statistics of the $T=0$ ensembles is much better, one observes an upward movement of data points as one goes from coarser to finer lattices, i.e. from  $n_{\tau}=4$ to $n_{\tau}$=8 (note the log-scale).

This enhancement might be indicative of a divergence due to mixing with lower-dimensional operators. 
With rare exceptions higher-dimensional operators suffer linearly divergent mixing 
with lower-dimensional operators in lattice regularization~\cite{Gockeler:1996mu}. 
In our setup these persist for $\hat{O}_{n},~n\ge1$ after vacuum subtraction [Eq.~\eqref{eq:qhatLatticeEquation}] due to temperature dependence of the lower-dimensional operators.
Both higher-twist operators also increase as the temperature is reduced; however, once the powers of temperature in the prefactors are taken into account, $T^2\hat{O}_{1}$ and $T^4\hat{O}_{2}$ decrease instead.

Whether the bump  at low temperatures  could be a signature of sensitivity to critical behaviour near the transition region is an open question. 
The full QCD result exhibits a similar pattern, albeit with large errors for $n_\tau=8$. Since we have not worked out the mixing of $\hat{O}_{1}$ and $\hat{O}_{2}$ operators with the respective lower{\color{red}-}dimensional operators, we estimate $\hat{q}$ in the limit of the hard parton energy $q^{-} \!\rightarrow \infty$, where the higher-twist terms do not contribute at all. 
 Given this restriction to the leading-twist operator, we switch, in the following, to the more suggestive notation 
\begin{equation}
\hat{O}_{0} \equiv \frac{FF}{T^4} \equiv \frac{\mathrm{Tr}[ F_{3j}F_{3j} - F_{4j} F_{4j}]}{T^4}.
\end{equation}

\subsection{Renormalization in Quenched QCD}
The expression for $\hat{q}$ [Eq.~\eqref{eq:qhatLatticeEquation}] 
applies after appropriate renormalization of the coupling and the field strength 
operators, for the hard quark traversing either the pure glue plasma, or a QGP. 
While the field strength operators mix in QCD with corresponding sea quark operators,
the latter do not contribute in Eq.~\eqref{eq:qhatLatticeEquation}, besides this mixing.
In quenched QCD, the renormalized leading-twist operator $\sfrac{FF}{T^4}$ is trivially 
related to  components of the renormalized EMT, here in the triplet representation $T^{(3)}$. 
The same relation holds for the bare variables: 
$FF \equiv [FF]^B = T_FT^{(3)B}$ (with $T_F=\sfrac{1}{2}$). 
Both undergo multiplicative renormalization with a (finite) factor $Z^{(3)}_{T}$ 
[Eq.~\eqref{eq:ZT3}] fixed by finite-momentum Ward identities, i.e.  
\bea
T^{(3)R} = Z^{(3)}_{T} T^{(3)B}, \text{ } \text{} 
\mathrm{and} \text{ } [FF]^R = Z^{(3)}_{T} [FF]^B. 
\eea
While $Z^{(3)}_{T}=1$ is trivial in the continuum theory, it explicitly depends 
on the particular discretization of the gauge field operator (in our case, a 
separate clover-leaf operator for each field strength tensor) and of the lattice 
gauge action that determines the background gauge field.
%
For the combination of the clover-leaf operator [Eq.~\eqref{eq:clover}] and Wilson's plaquette 
action this renormalization factor has been obtained in pure gauge theory~\cite{Giusti:2014ila,Giusti:2015daa,Giusti:2016iqr} 
using the shifted boundary condition approach~\cite{Giusti:2010bb,Giusti:2011kt,Giusti:2012yj}. 
This approach also 
carries over to QCD with sea quarks.
\begin{figure}[h!]
  \centering
    \includegraphics[width=0.8\linewidth]{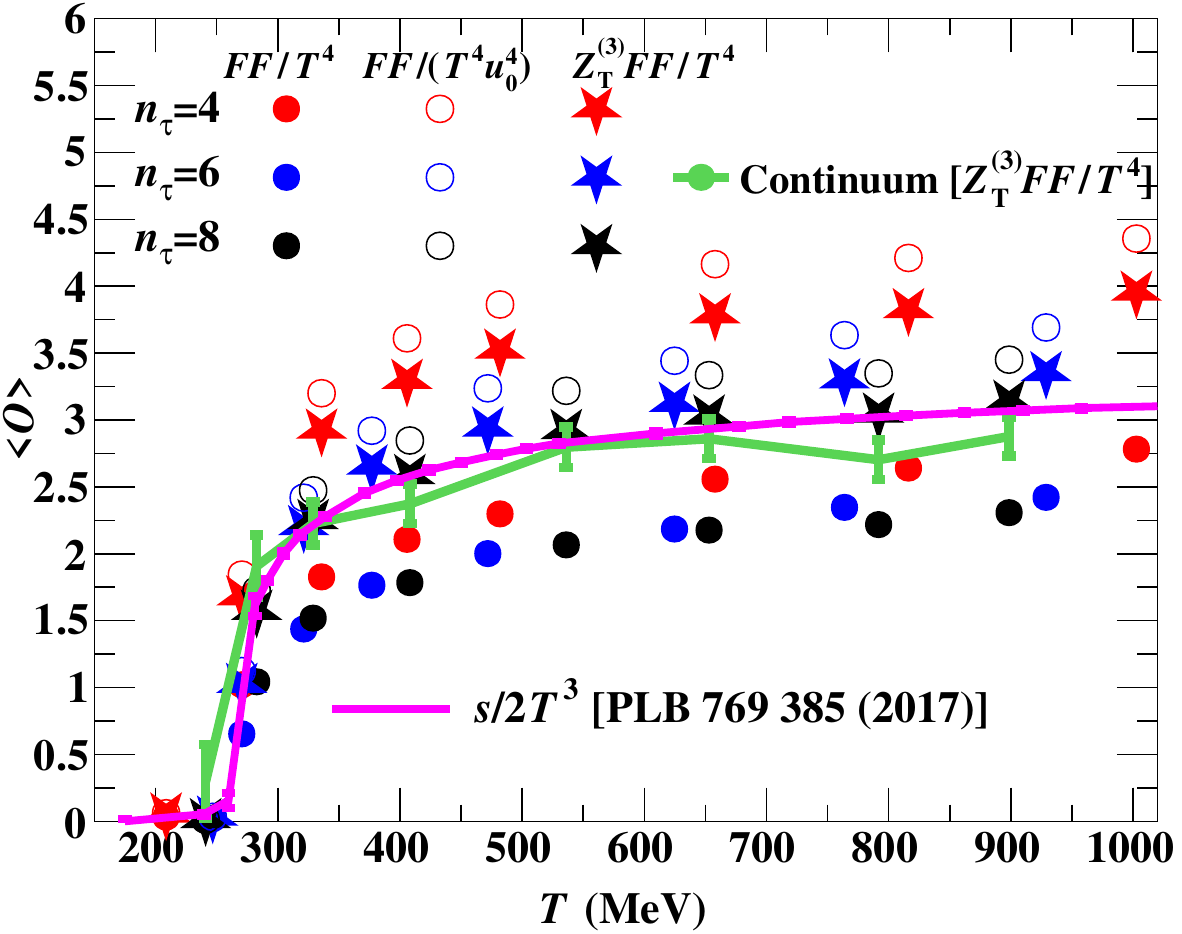}
  \caption{
  Bare and renormalized leading-twist operator 
  $\langle \mathrm Tr[F^{3i}F^{3i}\!-\!F^{4i}F^{4i}]\rangle/T^4$ in pure SU(3) gauge theory. 
  Our result agrees with the entropy density obtained in shifted boundary condition 
  approach~\cite{Giusti:2016iqr} after rescaling by $T_F=\sfrac{1}{2}$.}
  \label{fig:TDependenceVacuumSubtractedOperators}
\end{figure}

Using published data~\cite{Giusti:2015daa} we renormalize (in $\overline{MS}$~\cite{Capitani:2002mp}) 
$\sfrac{FF}{T^4}$ in pure gauge theory by converting the sextet renormalization 
factor $Z_T^{(6)}(g_0^2)$~\cite{Giusti:2016iqr} (clover-leaf) to 
triplet via 
\bea
Z_T^{(3)}(g_0^2) = z_T(g_0^2) Z_T^{(6)}(g_0^2).
\eea
The $Z^{(3)}_{T}$ renormalization factor is given by 
\begin{equation}
\begin{split}
& Z^{(3)}_{T} \!\!= \!\! \left[ \frac{1-0.509g^{2}_{0}}{1-0.4789g^{2}_{0}}  \right] \\
& \!\times \! \left[ \frac{1\!-\!0.4367g^{2}_{0}}{1\!-\!0.7074g^{2}_{0}}  
\!-\!0.0971g^{4}_{0} \!+\! 0.0886 g^{6}_{0} \!-\! 0.2909 g^{8}_{0}
\right].
\end{split}
\label{eq:ZT3}
\end{equation}
The renormalization factor has an error of up to 1\% for $g_0 \le 1$. 
While its error for $g_0 > 1$ is not known, it is certainly larger.
We account for it indirectly when performing the continuum extrapolation.
We interpolate $\sfrac{Z_T^{(3)}FF}{T^4}$ on the coarser ensembles 
($n_\tau \le 6$) linearly to the temperatures of the finest ensemble ($n_\tau = 8$), 
and then extrapolate at each temperature the two finest ensembles linearly 
[$\propto \sfrac{1}{n_\tau^2}$] or all three ensembles with a further quadratic term 
[$\propto \sfrac{1}{n_\tau^4}$] to the continuum. 
%
%
The linear fit provides the central value and the statistical error, while the spread between the central value from the linear fit and the quadratic fit give us the systematic error. Both the errors are added in quadrature and shown in green vertical bar in Fig.~\ref{fig:TDependenceVacuumSubtractedOperators}.
Our results agree with the $T_F$-rescaled entropy density using the shifted boundary 
condition approach~\cite{Giusti:2016iqr}, while estimating $Z^{(3)}_T(g_0^2)$ 
as $\sfrac{1}{u_0^4(g_0^2)}$ 
-- with tadpole factor 
\bea
u_0(g_0^2) = \sqrt[4]{\frac{\langle \mathrm{Tr} [U_{\mu,\nu}] \rangle}{N_c}}
\eea
-- yields roughly 10\% higher values (Fig.~\ref{fig:TDependenceVacuumSubtractedOperators}).

\subsection{Renormalization of leading-twist operator in full QCD }

The calculation of  $\hat{q}$ is substantially more involved in QCD than in the quenched approximation. 
In pure gauge theory, on the one hand, the renormalized leading-twist operator 
$\sfrac{FF}{T^4}$ is a genuine observable that is trivially related 
to the triplet component $T^{(3)}$ of the renormalized EMT.  
In the rest frame, the EMT's triplet component coincides with the 
entropy density times the temperature, $T^{(3)R} = s T$, underscoring the 
status of $[FF]^R$ as a scheme-independent observable in the pure gauge theory. 
In QCD, on the other hand, the leading-twist operator is not scheme-independent, 
and the previous relation to the entropy density $s$ does not hold. Instead, 
the renormalized leading-twist operator satisfies $[FF]^R = T_FT^{(3)R}_{G}$, 
i.e. only the renormalized gauge field operator's contribution to the EMT is 
considered, while the gauge background and higher order terms contain 
explicit contributions from the quark sea. 
The full entropy density $s$ is indeed a scheme-independent observable, and its 
renormalization is fixed by finite-momentum Ward identities, i.e. 
\bea
s T = T^{(3)R}_{G+Q} = Z^{(3)}_{G} T^{(3)B}_{G} + Z^{(3)}_{Q} T^{(3)B}_{Q}
\eea
in the rest frame.  
Both $Z^{(3)}_{G}$ and $Z^{(3)}_{Q}$ are finite, and can be fixed by two 
different finite momentum Ward identities using two different values of imaginary chemical potential. 
Here, $T^{(3)B}_{G} = [FF]^B$ is the same bare gauge field operator as in pure gauge 
theory (but on a QCD background), while $T^{(3)B}_{Q}$ is its valence quark 
counterpart ($N_f$ explicit contributions, i.e. from each of the sea quarks). 
We note that the choice of the regularization of $T^{(3)B}_{Q}$ does not have 
to coincide with the choice of the quark action of the QCD background fields.
In lattice-regularized QCD, the renormalized gauge field and quark operators 
are related to the bare ones by a mixing matrix $\mathcal{Z}$ as 
\be
\bvec T^{(3)R}_{G} \\ T^{(3)R}_{Q} \evec 
= \mathcal{Z} \bvec T^{(3)B}_{G} \\ T^{(3)B}_{Q} \evec, \quad 
\mathcal{Z} \equiv \bmat{cc} \mathcal{Z}^{(3)}_{GG} & \mathcal{Z}^{(3)}_{GQ} \\
   \mathcal{Z}^{(3)}_{QG} & \mathcal{Z}^{(3)}_{QQ} \emat 
\ee
where $\mathcal{Z}^{(3)}_{GG} \equiv Z^{(3)}_{G} + z_G$ 
and $\mathcal{Z}^{(3)}_{QQ} \equiv Z^{(3)}_{Q}+ z_Q$. 
The off-diagonal components $\mathcal{Z}^{(3)}_{GQ} \equiv -z_Q$ or 
$\mathcal{Z}^{(3)}_{QG}\equiv -z_G$ diverge as the regulator is removed, 
and so do the bare operators. 
Moreover, the coefficients $z_{G,Q}$ cannot be fixed using Ward identities, 
such that additional renormalization conditions need to be chosen to fix these 
in some particular scheme. 
Hence, $T^{(3)R}_{G}$ (and $T^{(3)R}_{Q}$) are renormalization scheme 
dependent in QCD. 
Without such a scheme being fixed before the regulator is removed, 
or without including the bare quark operator $T^{(3)B}_{Q}$, $T^{(3)R}_{G}$ 
and its continuum limit cannot be defined in QCD at all. 
For our lattice setup in QCD, neither the renormalization factors are known, 
nor the bare quark operators have been computed. 

For these reasons we currently can only produce an estimate of the renormalized 
leading-twist 
operator $[FF]^R$ for (2+1)-flavor QCD based on various complementary arguments. 
The first set of such considerations are of a purely quantitative nature, and 
concern reasonable estimates of the non-perturbative renormalization factors and 
cutoff effects in pure gauge theory that are transferred to the full QCD case. 
In pure gauge theory, the $\sfrac{FF}{T^4}$ with tadpole factor yields 
a 10\% shift from the renormalized result. 
Also, the magnitude of the bare $\sfrac{FF}{T^4}$ for $n_\tau\!\!=\!6$ is about 
10\% higher than the $n_\tau\!\!=\!8$ 
due to the cutoff effects in pure gauge theory. A similar trend is observed for 
$\sfrac{Z_T^{(3)}FF}{T^4}$, when comparing to the continuum limit. 
For full QCD case, based on 1-loop considerations (see below), we estimate that 
mixing with quark operators, not accounted for, may be at most an effect of 
commensurate size. 
Thus, we expect a deviation of no more than 30\% (adding all three sources of 
systematic uncertainty) between 
 $FF/(T^4u^4_{0})$ [$n_\tau\!\!=\!6$] and the correctly renormalized continuum limit. 

The second set of considerations relies on properties of the equation of state. 
The nonperturbative entropy density $s(T)$ or pressure $P(T)$~\cite{Bazavov:2017dsy,Weber:2018bam} 
are about 30\% below the Stefan-Boltzmann limit at $T \sim 2\,T_{pc}$, with the 
deviation diminishing by almost half at $T \sim 1$ GeV. 
For these and higher temperatures, the nonperturbative results are bracketed by 
electrostatic QCD (EQCD) at $\mathcal{O}(g^6)$~\cite{Laine:2006cp} and HTL-resummed 
perturbation theory at 3-loop order~\cite{Haque:2014rua} with less than 10\% deviation. 
This justifies assuming that the relative size of the (renormalized) gluon fraction 
of the full nonperturbative result can be estimated in the weak-coupling limit, 
i.e. the gluon fraction of $s(T)$ (and thus $T^{(3)}$) in (2+1)-flavor QCD being 
approximately $R_\mathrm{SB}~=~\sfrac{32}{95} \approx 0.337$ (of the SB limit). 
Thus, scaling down $T_F s T$ by this factor we may arrive at a QCD estimate of the
renormalized leading-twist operator $[FF]^R$ that is quantitatively similar to 
the previous estimate.

\begin{figure}[h!]
  \centering
    \includegraphics[width=0.8\linewidth]{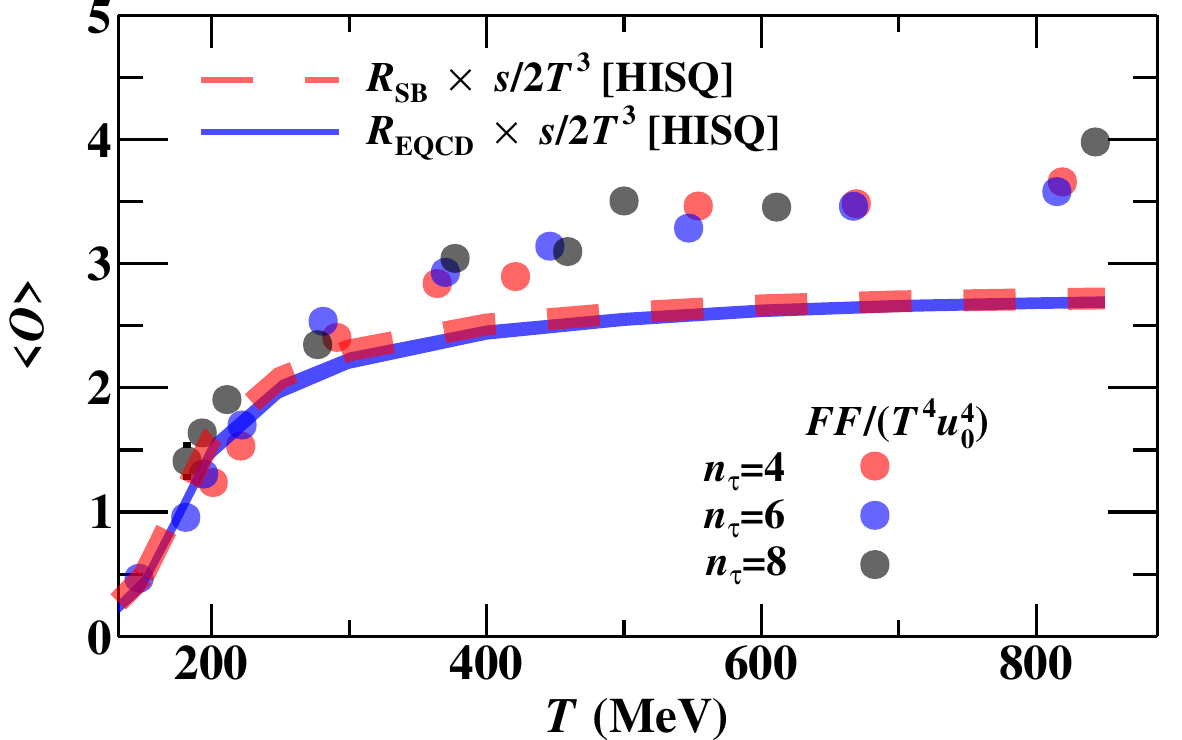}
  \caption{
  Comparison of the tadpole-factor rescaled leading-twist operator 
  $\langle \mathrm{Tr}[F^{3j}F^{3j} - F^{4j}F^{4j} ]/(T^4 u^{4}_{0}) \rangle$ 
  with $T_F$-rescaled entropy $\sfrac{s}{T^{3}}$ in (2+1)-flavor QCD~\cite{Bazavov:2017dsy,Weber:2018bam} 
  scaled by the gluon fraction in the Stefan-Boltzmann limit ($R_\mathrm{SB}$) or in EQCD ($R_\mathrm{EQCD}$)~\cite{Laine:2006cp}.
  }
  \label{fig:EstimatesEQCD_HISQ_FFoperators}
\end{figure}

The aforementioned spread of up to 30\% between  the non-perturbative result and 
the SB limit appears to be a fairly cautious estimate of the uncertainty 
associated with this estimate of $[FF]^R$ for $T \gtrsim 2\,T_{pc}$.
Defining the ratio $R_\mathrm{EQCD}(\sfrac{T}{T_{pc}})$ between the $N_f=0$ to 
$N_f=3$ EQCD results at $\mathcal{O}(g^6)$ and fixed value of 
$\sfrac{T}{T_{pc}}$, and rescaling the (2+1)-flavor QCD lattice result 
is expected to be an even better estimate, 
since the EQCD results are even more similar to the lattice data. 
In Fig.~\ref{fig:EstimatesEQCD_HISQ_FFoperators} we show that both estimates 
of the gluon fraction of the entropy density in full QCD yield 
similar results that are within the 30\% uncertainty margin, and confirm 
the expectation of a downward correction for the continuum limit.

Alternatively, we may take a closer look at an instance of the mixing matrix 
$\mathcal{Z}$ in QCD, which is known -- for some particular set of discretized 
gauge field and quark operators, with QCD background fields in terms of  
some particular combination of lattice actions -- at the 1-loop level.
Its $N_f$-independent coefficients $Z_{G,Q}^{(3)}$ at $\mathcal{O}(g_0^2)$ are 
one order of magnitude larger [$\sim 10\%$] ($\times N_c$) than the $N_f$-dependent 
ones [$\sim 1\%$] ($\times N_f$) for the combination of Wilson's plaquette action 
and (unimproved) Wilson fermion action~\cite{Capitani:1994qn,DallaBrida:2020gux}. 
Similar statements (in terms of magnitudes) hold for the coefficients $z_{G,Q}$ 
at the 1-loop level~(and typical couplings $g_0 \!\sim\!1$). 
The magnitudes of the coefficients change somewhat with the discretization, 
e.g. the $N_f$ dependent 1-loop coefficients change within a factor $4$ 
between unimproved or improved Wilson fermions~\cite{Capitani:1994qn,DallaBrida:2020gux}; 
for improved Wilson fermions the sum of the $N_f$-dependent coefficients is 
as large as the sum of the $N_f$-independent ones. 
As there is no obvious reason why the magnitudes of such coefficients should not 
be similar for the combination of discretizations in our case, i.e. HISQ action 
and Symanzik gauge action, we anticipate that these findings apply within 
a factor $2$ to our combination of tree-level Symanzik gauge and HISQ action.

The smallness of the $N_f$-dependent coefficients (both for unimproved or improved 
Wilson fermions) at $\mathcal{O}(g_0^2)$ suggests that the error (for any given 
bare coupling $g_0$) due to neglecting the mixing with quark contributions from 
$N_f$ light flavors is below $10\%$, and while use of a multiplicative 
renormalization factor for $\sfrac{FF}{T^4}$ based on a different gauge action may 
be off at the $30\%$ ($\approx 10\% \times N_c$) level, which constitutes (for any 
given bare coupling $g_0$) the quantitatively dominant uncertainty. 
Since bare $[FF]^B$ operators at similar $\sfrac{T}{T_{pc}} \gtrsim 2$ (i.e. 
$T \approx 300$ MeV in (2+1)-flavor QCD corresponding to $T \approx 500$ MeV in 
pure gauge theory) are within $20\%$ of each other for the pure gauge and full QCD 
ensembles despite their major differences (different background fields and choices 
of the lattice action), the uncertainty related to the renormalization factor may 
be considered as dominant.

Concluding this line of reasoning, one might expect that we could also multiply the bare 
$[FF]^B(g_0^2)$ determined in (2+1)-flavor QCD by $Z_T^{(3)}(g_0^2)$ (determined 
in pure gauge theory) and obtain yet another estimate, which is quantitatively 
similar to the first one. 
However, this approach does not work in practice, since the pure 
gauge theory parameterization of the renormalization factor for the Wilson 
plaquette action has a pole in the middle of the range of bare gauge coupling 
for the Symanzik gauge action used in the (2+1)-flavor QCD simulations.
Nevertheless, based on these 1-loop considerations we expect that an overall 30\% 
uncertainty is justified as a reasonably cautious assessment for our 
estimate of $[FF]^R$ in full QCD.

\section{Results}
\label{sec:results}

\begin{figure}[h!]
\vspace{-0.25cm}
\includegraphics[width=0.92\linewidth]{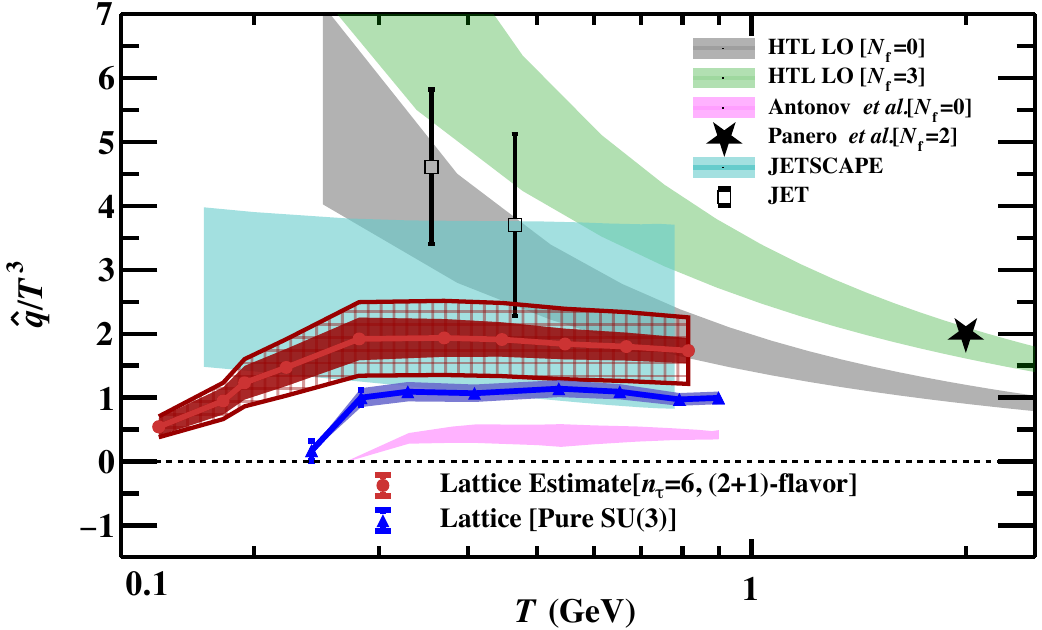}
\vspace{-0.25cm}
\caption{ Lattice determination of $\hat{q}$ for a highly energetic hard 
quark traversing a pure glue (blue) and (2+1)-flavor QCD (red) plasma. Error bars show statistical errors, solid bands represent scale variation.
The pure SU(3) result is renormalized and continuum extrapolated. 
For the QCD estimate, systematic errors due to estimated renormalization (via $u_0$) and lattice cutoff ($n_\tau\!=\!6$) are indicated by the checkered band.
Also plotted are the phenomenological extractions from the JET~\cite{Burke:2013yra} and JETSCAPE~\cite{Soltz:2019aea} collaborations, 
non-perturbative results from an $N_f\!=\!2$, 3-D lattice calculation~\cite{Panero:2013pla}, 
an $N_f\!=\!0$ stochastic vacuum model calculation~\cite{Antonov:2007sh}, 
and LO-HTL calculations from Ref.~\cite{He:2015pra}, with $\alpha_S$  evaluated at $2\pi T \leq \mu \leq 4\pi T$.}
  \label{fig:LatticeQhatResult}
\end{figure}

In Fig.~\ref{fig:LatticeQhatResult} we present the resulting $\hat{q}/T^3$ based on Eq.~\eqref{eq:qhatLatticeEquation}. The coupling at the vertex to gluons absorbed by the medium must be at the temperature scale. 
We vary this scale as $(2 - 4)\pi T$ to account for the truncation error and  use of the  1-loop gauge coupling.
While the non-perturbative renormalization factors for the (2+1)-flavor QCD 
result are unknown, we have used several means to 
obtain well-justified estimates. 
%
As we expect a deviation of no more than 30\% 
between $FF/(T^4u^4_{0})$ [$n_\tau\!\!=\!6$] and the correctly renormalized continuum limit,
we attach a symmetric relative uncertainty of 30\% to this lattice QCD estimate. 
We multiply by the  1-loop gauge coupling (same scale variation) for $N_f=3$.

Due to the OGE approximation~[Eqs.~\eqref{eq:qhatDefinitionBoltzmann},\eqref{eq:OGE-qhat}], 
truncation at leading twist [Eq.~\eqref{eq:qhatLatticeEquation}], 
and the coupling $g(T)$ at the temperature scale, $q^{-}$ dependence is 
absent in our result for $\sfrac{\hat{q}}{T^3}$.
Hence, this result applies in the limit $q^{-} \!\to \infty$ of an infinitely hard parton.
The temperature dependence of the resulting $\sfrac{\hat{q}}{T^3}$ is 
shown in Fig.~\ref{fig:LatticeQhatResult} for the continuum limit of  pure SU(3) gauge theory (blue) or for our estimate in (2+1)-flavor 
lattice QCD (red). 

The transport coefficient $\sfrac{\hat{q}}{T^{3}}$ exhibits a rapid rise in the transition region and slightly above, i.e. in the temperature 
range $150\,\mathrm{MeV} \! \lesssim \! T \! \lesssim \! 250\,\mathrm{MeV}$ for 
(2+1)-flavor QCD or $250\,\mathrm{MeV} \! \lesssim \! T \! \lesssim \! 350\,\mathrm{MeV}$ 
for the pure SU(3) gauge theory, and is flat within errors above 
$400\,\mathrm{MeV}$. 
The change of the gauge coupling $g(T)$ with $T$ partially compensates 
the temperature dependence of the leading-twist operator at temperatures well above $T_{pc}$.
Interestingly, the nonperturbative stochastic vacuum model 
result~\cite{Antonov:2007sh} exhibits a very similar behavior.

Expectedly, the lattice results do not show any log-like rise at lower $T$, 
as one observes in leading-order (LO) HTL calculations~\cite{Qin:2009gw}
(for the HTL bands in Fig.~\ref{fig:LatticeQhatResult} $q^-\!\!=\!100$ GeV is assumed). 
This arises from the dominant diagram with 
OGE in the perturbative 
calculation, which leads to a logarithm in $q^{-}/T$. 
Interestingly, no such logarithm arises at next-to-leading-order (NLO) in 
the HTL expansion of $\hat{q}$~\cite{Caron-Huot:2008zna}. 
The finite part of the NLO result is much larger than the LO result and far above the scale in Fig.~\ref{fig:LatticeQhatResult}.
Similar contributions may appear 
once some approximations used in this paper are lifted, e.g. if emission 
of gluons is considered, or if the higher-twist operators become 
non-negligible as $q^-\!\to\!\infty$ is relaxed. 
Whether such terms will dominate remains to be determined. 
The \mbox{3-D} lattice simulation~\cite{Panero:2013pla} exhibits a behavior quite similar to perturbative HTL; the result at $T=400\,\mathrm{MeV}$ is far above the scale in Fig.~\ref{fig:LatticeQhatResult}.

In Fig.~\ref{fig:LatticeQhatResult}, we also present a comparison with 
phenomenological extractions of $\sfrac{\hat{q}}{T^3}$ 
obtained by the JET~\cite{Burke:2013yra} and JETSCAPE~\cite{Soltz:2019aea} collaborations. 
The JET collaboration applied several disparate models of energy loss with 
either a sole $T$ dependence of the ratio $\sfrac{\hat{q}}{T^3}$, or one 
obtained from HTL effective theory~\cite{Caron-Huot:2008zna}. 
The JETSCAPE extraction applied an amalgam of theories for different epochs 
of the jet shower, with a data-driven (Bayesian) determination of 
$\sfrac{\hat{q}}{T^3}$, allowed to depend on $T$, the energy and scale of a 
given parton in the shower. 
A log-like rise at low $T$ is allowed in both frameworks; both work with the OGE approximation.

\section{Conclusion}
\label{sec:conclusions}

In this paper, we carried out the first rigorous \emph{first-principles} 
\mbox{4-D}~calculation of the jet quenching parameter $\hat{q}$, which is the leading 
coefficient affecting jet modification in the QGP. 
We computed $\hat{q}$~for a single parton 
undergoing a single scattering off the medium, utilizing lattice gauge 
theory in the quenched approximation. We outlined the specific challenges 
of a corresponding (2+1)-flavor lattice calculation, while providing a first 
theoretically motivated lattice estimate of $\hat{q}$ in (2+1)-flavor QCD. 

While the proximity of the lattice calculations with phenomenological 
extractions is very encouraging, several caveats need to be considered: 
The full QCD result is only an estimate, due to lack of rigorous control 
of the renormalization factors and the mixing with still unknown quark 
operators on the lattice. 

As the $q^- \to \infty$ limit is relaxed, the perturbative portions of the current calculation will have to be extended 
to higher-order, allowing for multiple scattering and emission in the medium. 
While we do not expect multiple scattering to yield contributions that cannot be factorized into independent scatterings 
(as is the case in all pQCD based jet quenching calculations and phenomenology, including the extractions from the JET and JETSCAPE collaborations),
emissions in the process of scattering may lead to shifts (in $\sfrac{\hat{q}}{T^3}$) of the order of the width of the bands in our QCD estimate\st{s}. As discussed in the Appendix~\ref{subs:RadiativeFactors}, applying known perturbatively calculated 
renormalization factors~\cite{Blaizot:2012fh,Blaizot:2014bha,Liou:2013qya}, will bring down the phenomenological extractions by about 33\%, dramatically increasing the agreement with our calculations. Future efforts which expand Eq.~\eqref{eq:qhatLatticeEquation} as a power series in $\sfrac{T}{q^-}$, will encounter mixings with novel quark operators at order $\sfrac{T}{q^-}$. At order $(\sfrac{T}{q^-})^2$, one will encounter mixing with possible linearly divergent, 
temperature dependent operators, that cannot be straightforwardly canceled via vacuum subtraction.

%
%
%

\emph{Acknowledgements.} 
We thank A. Patella, R. Sommer, and M. Dalla Brida for extensive discussion about the renormalization of the EMT. 
This work was supported in part by the National Science Foundation (NSF) under grant number~{ACI-1550300} (JETSCAPE), 
by the U.S. Department of Energy (DOE) under grant number~{DE-SC0013460}.
JHW's research is funded by the Deutsche Forschungsgemeinschaft (DFG, 
German Research Foundation) - Projektnummer 417533893/GRK2575 ``Rethinking 
Quantum Field Theory''. The calculations were performed using High Performance Computing (HPC) facility at Wayne State University funded by the Wayne State OVPR. The data storage was provided in part by
the OSIRIS project supported by the National Science
Foundation under grant number OAC-1541335.

\begin{appendices}
\section{$\mbox{}$\!\!\!\!\!\!: Gauge ensembles and lattice setup}
\label{sec:GaugeEnsemble}

In this section, we list the parameters used in generating gauge ensembles for pure SU(3) and (2+1)-flavor QCD lattices. In the presented lattice calculations, the unquenched lattices were generated at the physical value of the strange quark mass $m_{s}$ and the light sea quark masses of $m_{l}=m_{s}/20$ using the HISQ~\cite{Follana_2007} and tree-level Symanzik improved gauge action~\cite{Bazavov:2014pvz,Bazavov:2018wmo}. 
We employed the Rational Hybrid Monte Carlo algorithm (RHMC)~\cite{Clark:2004cp}. 
In Table \ref{table:FullQCDNt4},  \ref{table:FullQCDNt6} and \ref{table:FullQCDNt8}, we present the  strange quark mass ($am_{s}$) in units of lattice spacing $a$, temperature ($T$) and time units (TU) for $n_{\tau}=4,6,8$ and their vacuum analog $T=0$. The temperatures for different $\beta_{0}=10/g^{2}_{0}$'s have been fixed using the $r_1$ scale and taken from Refs. \cite{Bazavov:2018wmo}. 

In Table \ref{table:PureSU3Nt4}, \ref{table:PureSU3Nt6}  and \ref{table:PureSU3Nt8}, we provide  $\beta_{0}$, temperature and the collected statistics  for pure SU(3) lattices. 
The scale setting was done using the two-loop perturbative renormalization group (RG) equation with non-perturbative correction factor [$f(\beta_{0})$]  given as 
\begin{equation}
a = \frac{f(\beta_{0})}{\Lambda_{L}}   \left [  \frac{ 11g^{2}_{0}} {16\pi^{2}} \right]^{\frac{-51}{121}}  \exp \left[   \frac{-8\pi^{2}}{11g^{2}_{0}}  \right]
\end{equation}
where $\Lambda_{L}$ is a lattice parameter. 
We estimated the non-perturbative  factor by adjusting the function $f(\beta_{0})$ such that $T_{c} /  \Lambda_{L}$ is independent of bare coupling constant $g_{0}$.
 In this calculation, $\Lambda_{L}$ was set $5.5$ MeV \cite{Hasenfratz:1980kn,Booth:2001qp,Deur_2016} and the critical temperature to $T_{c}\approx 265$ MeV \cite{Philipsen:2012nu}.

\begin{center}
\begin{table}[h!]
 \begin{tabular}{| c c c c  c |} 
 \hline \hline
 $\beta_{0}=10/g^{2}_{0}$     &         $am_{s}$     &     $T$ (MeV)      &     $\#$TUs($T$$\ne$0) & $\#$TUs($T$=0) \\ [0.98ex] 
 \hline
5.9       &     0.132     & 		201   &		10000	   &  10000				\\
6.0       &     0.1138  	&			221	  &		10000		&	10000		\\
6.285  &      0.079   	&			291	  &		10000	    &  10000		\\
6.515   &     0.0603  &			364	  & 		10000	    &  10000	   \\
6.664  &     0.0514  	&			421	  &		20000	&   10000	  \\
6.95     &     0.0386 &			554	  &		10000	    &     10000		  \\
7.15      &     0.032   	&			669  &	    10000     &     10000   	 \\
7.373   &     0.025   	&			819   &		 10000    &   10000    \\
\hline \hline
\end{tabular}
\caption{The parameters to generate (2+1)-flavor QCD gauge ensembles with  $m_{l}=m_{s}/20$ for lattice size $n_{\tau}=4$ with aspect ratio $n_{s}/n_{\tau}=4$ .}
\label{table:FullQCDNt4}
\end{table}
\end{center}
%
%
%
\begin{center}
\begin{table}[h!]
 \begin{tabular}{||c c c c  c||} 
 \hline \hline
 $\beta_{0}=10/g^{2}_{0}$     &         $am_{s}$     &     $T$ (MeV)         &     $\#$TUs($T$$\ne$0) & $\#$TUs($T$=0)  \\ [0.98ex] 
 \hline
6.0     &	0.1138  &	147  & 	 	10000 	& 	10000 \\
6.215   &	0.0862  &	181  & 	 	10000   & 	10000  \\
6.285   &	0.079   &	194  & 	 	10000   & 	10000 \\
6.423   &	0.067   &	222  & 	    7600	& 	10000 \\
6.664   &	0.0514  &	281  & 	 	10000	& 	 7000  \\ 
6.95    &	0.0386  &	370  & 	   10000 	& 	 8000 \\
7.15    &	0.032   &	446  & 	 	10000	& 	 8600 \\
7.373   &	0.025   &	547  & 	   10000	& 	10000 \\
7.596   &	0.0202  &	667  & 	    8600 	& 	10000 \\
7.825   &	0.0164  &	815  & 	    9140 	& 	10000 \\
\hline \hline
\end{tabular}
\caption{The parameters to generate (2+1)-flavor QCD gauge ensembles with  $m_{l}=m_{s}/20$ for lattice size $n_{\tau}=6$ with aspect ratio $n_{s}/n_{\tau}=4$ .}
\label{table:FullQCDNt6}
\end{table}
\end{center}
\begin{center}
\begin{table}[h!]
 \begin{tabular}{||c c c c  c||} 
 \hline \hline
 $\beta_{0}=10/g^{2}_{0}$     &         $am_{s}$     &     $T$ (MeV)      &     $\#$TUs($T$$\ne$0) & $\#$TUs($T$=0)   \\ [0.98ex] 
 \hline
6.515      &       0.0604 &   	182  &  	 7300  &  	 6400    \\
6.575      &       0.0564 &   	193    &      8650  &      6800    \\
6.664      &       0.0514 &   	211    &     10000   &  	 5000 	 \\
6.95       &       0.0386 &   	277    &  	10000  &  	 5950 \\
7.28       &       0.0284 &   	377    &     10000     &   6550  \\
7.5        &       0.0222 &   	459    &  	10000  &  	 5000	 \\
7.596      &       0.0202 &   	500  &    	10000	  &   9400  \\ 
7.825      &       0.0164 &   	611   &   	10000	  &   7900  \\
8.2        &       0.01167 &  	843   &   	10000	  &   5000  \\
\hline \hline
\end{tabular}
\caption{The parameters to generate (2+1)-flavor QCD gauge ensembles with  $m_{l}=m_{s}/20$ for lattice size $n_{\tau}=8$ with aspect ratio $n_{s}/n_{\tau}=4$ .}
\label{table:FullQCDNt8}
\end{table}
\end{center}
%
%
%

\begin{center}
\begin{table}[h!]
 \begin{tabular}{||c c  c  c||} 
 \hline \hline
 $\beta_{0}=6/g^{2}_{0}$        &     $T$ (MeV)      &     $\#$TUs($T$$\ne$0) & $\#$TUs($T$=0)  \\ [0.98ex] 
 \hline
 5.6       &   209    & 	10000    & 	10000  \\
5.7        &  271     & 	10000    & 	10000 \\ 
5.8        &  336      & 	10000    & 	10000 \\
5.9       &   406	    & 10000    & 	10000 \\
6.0       &   482    & 	10000	    & 10000     \\
6.2       &   658    & 	10000	    & 10000     \\
6.35      &   816    & 	10000	    & 10000 \\
6.5       &   1003    & 	10000    & 	10000 \\
6.6       &   1146    & 	10000    & 	10000 \\
\hline \hline
\end{tabular}
\caption{The parameters to generate pure SU(3) gauge ensembles using Wilson's pure SU(3) gauge action for lattice size $n_{\tau}=4$ with aspect ratio $n_{s}/n_{\tau}=4$ .}
\label{table:PureSU3Nt4}
\end{table}
\end{center}
%
%
%
\begin{center}
\begin{table}[h!]
 \begin{tabular}{||c  c c  c||} 
 \hline \hline
$\beta_{0}=6/g^{2}_{0}$          &     $T$ (MeV)      &     $\#$TUs($T$$\ne$0) & $\#$TUs($T$=0)  \\ [0.98ex] 
 \hline
 5.60     &         139   &       	10000   &       	10000    \\   
5.85     &         247   &       	10000    &        	10000    \\   
5.90     &         271   &       	10000	   &       10000    \\   
6.00     &         321	   &       10000   &       	10000    \\   
6.10     &         377	   &       10000   &       	10000    \\   
6.25     &         472	   &       10000	   &       10000    \\   
6.45     &         625   &       	10000	   &       10000    \\   
6.60     &         764	   &       10000   &       	10000    \\   
6.75     &         929	   &       10000	   &       10000    \\   
6.85    &         1056	   &       10000	   &       10000    \\   
\hline \hline
\end{tabular}
\caption{The parameters to generate pure SU(3) gauge ensembles using Wilson's pure SU(3) gauge action for lattice size $n_{\tau}=6$ with aspect ratio $n_{s}/n_{\tau}=4$ .}
\label{table:PureSU3Nt6}
\end{table}
\end{center}
%
%
\begin{center}
\begin{table}[h!]
 \begin{tabular}{||c  c c  c||} 
 \hline \hline
$\beta_{0}=6/g^{2}_{0}$     &     $T$ (MeV)      &     $\#$TUs($T$$\ne$0) & $\#$TUs($T$=0)   \\ [0.98ex] 
 \hline
 5.70    &       135   &       10000   &      10000     \\   
5.95     &      221	  &    10000	  &    10000    \\   
6.00     &      241  &    	10000	  &    10000    \\   
6.10      &     283  &    	10000	  &    10000    \\   
6.20     &      329	  &    10000  &    	10000    \\   
6.35    &       408	  &    10000	  &    10000    \\   
6.55    &       536	  &    10000	  &    10000    \\   
6.70    &       653	  &    10000	  &    10000    \\   
6.85    &       792	  &    10000  &    	10000    \\   
6.95     &      899	  &    10000	  &    10000    \\   
\hline \hline
\end{tabular}
\caption{The parameters to generate pure SU(3) gauge ensembles using Wilson's pure SU(3) gauge action for lattice size $n_{\tau}=8$ with aspect ratio $n_{s}/n_{\tau}=4$ .}
\label{table:PureSU3Nt8}
\end{table}
\end{center}

\section{$\mbox{}$\!\!\!\!\!\!: Radiative renormalization factors}
\label{subs:RadiativeFactors}

\begin{figure}[h!]
  \centering
    \includegraphics[width=0.93\linewidth]{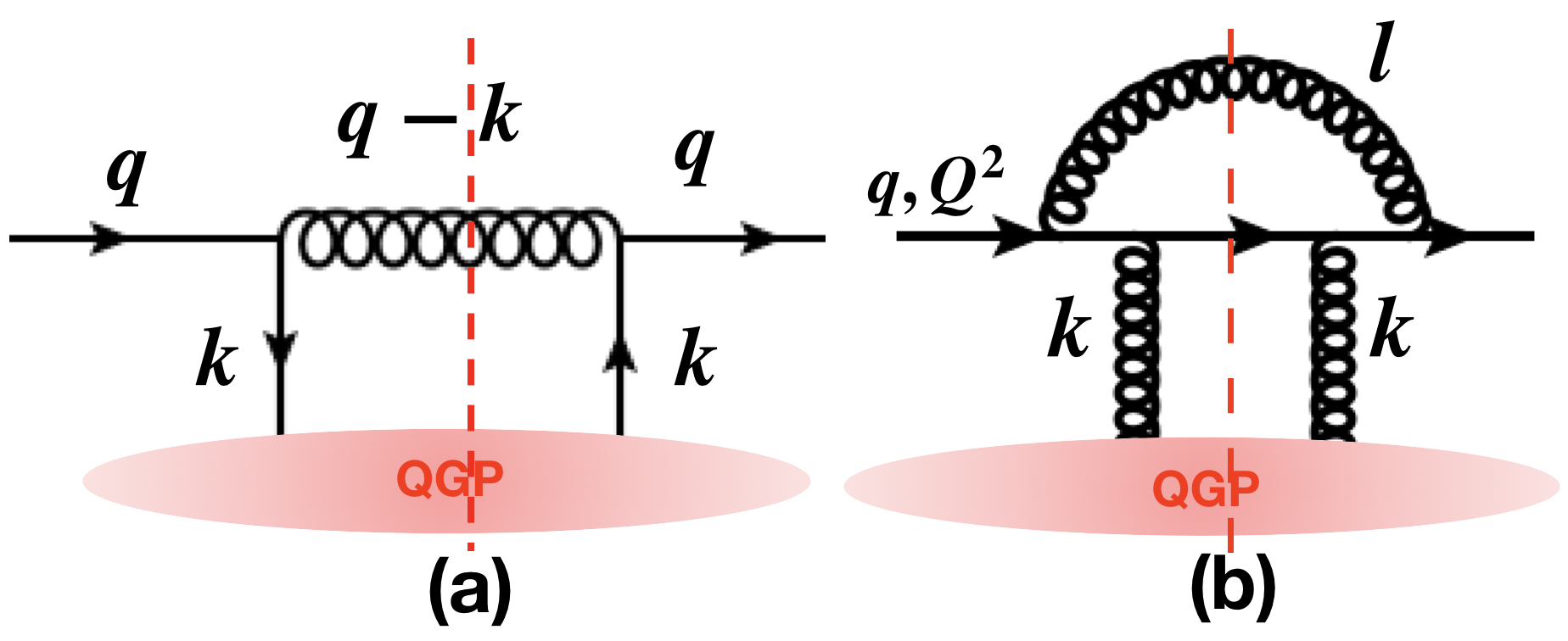}
  \caption{ Forward scattering diagrams for future outlook. (a) Transverse broadening due to flavor changing process. (b)  A typical single scattering and single emission diagram.   }
  \label{fig:NewDiagrams}
\end{figure}

As we relax the $q^- \rightarrow \infty$ limit, there are 4 categories of new contributions that will modify the results obtained in the current calculation. First are the quark operators which will mix with the gluon operators in the process of renormalization in (2+1)-flavor QCD. The determination of the magnitude and mixing with these operators is the next step for full QCD simulations for $\hat{q}$. While the magnitude, mixing, and eventual effect of these terms on \qhat are expected to be small, these terms may have other phenomenological effects on jets. 
Next is the appearance of higher twist terms. These have already been described above. 
The other contributions would include processes that involve flavor change [Fig.~\ref{fig:NewDiagrams}(a)], multiple scattering or emission inside the lattice itself [Fig.~\ref{fig:NewDiagrams}(b)]. 

Every calculation of jet modification, other than in AdS/CFT, assumes that multiple scattering can be factorized into multiple independent scatterings,
 and we don't expect a different result here. The more interesting case is the modification to the collision kernel due to radiative effects. 
 Of course, such calculations have never been carried out on the lattice. 
 However, these have been evaluated in continuum perturbation theory~\cite{Liou:2013qya,Blaizot:2012fh,Blaizot:2014bha}, for media whose length is shorter than the formation time of a radiation as,
 \begin{equation}
 \hat{q}^R = \hat{q} + \delta \hat{q} = \hat{q} \left[ 1 + \frac{\alpha_S N_c }{4 \pi} \log^2 \left(  \frac{L^2}{l_0^2} \right) \right].
 \end{equation}
  In the equation above, $L$ is the length of the medium and $l_0$ is the approximate size of a scattering center, which in a thermal medium is approximately the thermal wavelength. 
  Thus, $l_0 \sim 1/T$ and as a result $L/l_0 \sim 4$, for the lattices used in this paper. Using $\alpha_S \simeq 0.25$, we obtain the additional corrections to be $\delta \hat{q} \sim 0.5 \hat{q}$. 
  Thus the perturbatively corrected magnitude of the transport coefficient $\hat{q}$ engenders an approximate $50\%$ excess in the value of $\hat{q}$.  One should note that the above estimate hinges on the knowledge of the exact value of $\alpha_S$, as well as the numerical factor $n$ in the equality $l_0 = n/T$. While we have assumed $n=1$, this can easily vary up or down by about 100\%, as the exact size of a scattering center is not well defined in a QGP. Such variations can lead to noticeable shifts in our estimate of $\delta$\qhat. If $\delta$\qhat were to become comparable to \qhat then additional higher order terms neglected in the equation above will have to be considered. The reader is directed to Refs.~\cite{Liou:2013qya,Blaizot:2012fh,Blaizot:2014bha} for extended discussion on these issues.

 Accepting our estimate for $\delta$\qhat above, we should further clarify what this 50\% excess means and how it should be applied to the comparison plot in Fig.~[3 of paper]. Note that while $\hat{q}$ is defined as a transverse broadening coefficient, it is typically extracted from experimental 
data by comparing the energy lost by jets and leading hadrons, due to excess radiation caused by the transverse exchanges with the medium. The $\delta \hat{q}$ factor above, describes a perturbatively calculated shift that should be applied to $\hat{q}$ when it is extracted from energy loss calculations. Thus, this factor should be used to reduce the values of $\hat{q}$ extracted by the JET and JETSCAPE collaborations, which obtained $\hat{q}$ by comparing energy loss calculations to data (without the $\delta \hat{q}$ factor). This will bring the JET points and the JETSCAPE band in Fig.~[3 of paper] to about 66\% of their current values, in complete agreement with the lattice calculation, which measures $\hat{q}$ from transverse broadening, without any emissions.

 In a future calculation of $\hat{q}$ on the lattice, which will include emissions [Fig.~\ref{fig:NewDiagrams}(b)], we will likewise encounter a shift in the final measured value of $\hat{q}$, due to the larger phase space available for the 
transverse momentum exchange. A large portion of this will be perturbative, equivalent to the factor $\delta \hat{q}$ above, as we will continue to assume that the jet and its emissions can be treated perturbatively. It is possible that there will also be a small non-perturbative renormalization, which would be obtained by comparing with the $\hat{q}$ calculated in the current study (without emissions), identifying the excess $\delta \hat{q}$, and subtracting the perturbative correction from this.  

\end{appendices}

\bibliography{refs}
\end{document}